\documentclass[aps,pre,
reprint,
groupedaddress,amsmath]{revtex4-1}
\usepackage{graphicx,color,dcolumn,tikz,pgf}

\newcommand{\mub}{\mu_{\rm B}}
\newcommand{\vk}{\vec k}
\newcommand{\ve}{\vec e}

\newcommand{\vc}{\vec c}

\newcommand{\vj}{\vec j}
\newcommand{\vm}{\vec m}

\newcommand{\vq}{\vec q}
\newcommand{\vp}{\vec p}
\renewcommand{\vr}{\vec r}

\newcommand{\vv}{\vec v}

\newcommand{\vS}{\vec S}

\newcommand{\vsigma}{\mbox{\boldmath $\sigma$}}

\renewcommand{\vec}[1]{\mathbf{#1}}

\begin{document}


\title{Boltzmann approach to the longitudinal spin Seebeck effect}


\author{Rico Schmidt, Francis Wilken, Tamara S. Nunner, and Piet W. Brouwer}
\affiliation{Dahlem Center for Complex Quantum Systems and Fachbereich Physik, Freie Universit\"at Berlin, 14195 Berlin, Germany}


\date{\today}

\begin{abstract}
We develop a Boltzmann transport theory of coupled magnon-phonon transport in ferromagnetic insulators. The explicit treatment of the magnon-phonon coupling within the Boltzmann approach allows us to calculate the low-temperature magnetic-field dependence of the spin-Seebeck voltage. Within the Boltzmann theory we find that this magnetic field dependence shows similar features as found by Flebus {\it et al.} [Phys.\ Rev.\ B {\bf 95}, 144420 (2017)] for a strongly coupled magnon phonon system that forms magnon-polarons, and consistent with experimental findings in yttrium iron garnet by Kikkawa {\it et al.} [Phys.\ Rev.\ Lett.\ {\bf 117}, 207203 (2016)]. In addition to the anomalous magnetic-field dependence of the spin Seebeck effect, we also predict a dependence on the system size.
\end{abstract}


\maketitle


\section{Introduction}

The spin Seebeck effect (SSE) describes the generation of a spin current in a magnetic material in response to a temperature gradient applied across the sample \cite{Uchida-2008,Jaworski-2010,Uchida-2010,Bauer-2012}. The spin current can be transferred to a paramagnetic metal (NM) layer attached to the magnet and is then typically detected via the inverse spin Hall effect (ISHE) \cite{Rezende-2005,Saitoh-2006,Valenzuela-2006,Maekawa-2007}. Although the spin Seebeck effect was first observed in metals \cite{Uchida-2008}, it has also been reported for magnetic semiconductors \cite{Jaworski-2010} and magnetic insulators \cite{Uchida-2010}. While the SSE remains controversial in the transverse configuration, in which spin current and temperature gradient are perpendicular, due to possible effects from out-of-plane temperature gradients, the longitudinal SSE, in which the temperature gradient and the spin current are collinear, has been reproduced by many groups \cite{Uchida-2010-3,Jaworski-2011,Uchida-2011,Weiler-2012,Uchida-2012,Schreier-2013,Kikkawa-2013,Uchida-2013,Uchida-2014}. Recently, the spin Seebeck effect has also been observed in antiferromagnets \cite{Seki-2015,Wu-2016}.

Whereas a spin current in a metallic ferromagnet can be carried by both conduction electrons and spin waves, in a ferromagnetic insulator (FMI) the spin current of the SSE is carried exclusively by spin waves or, using a quantum-mechanical language, ``magnons''. At the same time, in an FMI the applied temperature gradient primarily affects the lattice vibrations, {\em i.e.}, the phonons. The initial theoretical works by Xiao {\it et al.} \cite{Xiao-2010} and Hoffman {\it et al.} \cite{Tserkovnyak-2013}, which treat the spin dynamics in the FMI in a Landau-Lifshitz-Gilbert approach, describe the effect of phonons on the magnetization dynamics by means of an effective temperature-dependent noise term. 
A second class of theoretical calculations by Rezende and co-workers \cite{Rezende-2014,Rezende-2014-2,Rezende-2016} is based on a Boltzmann approach. Whereas this approach tackles the role of magnon-magnon interaction to the SSE inside the FMI in great detail, it attributes the (phonon-related) thermal relaxation processes of the magnons in terms of a phenomenological thermal lifetime $\tau_{\rm mp}$. In both theories, the magnon-phonon interaction plays a key role in the determination of the magnon mean free path and, thus, of the system-size and the magnetic field dependence of the magnon-driven SSE. A purely phenomenological treatment of the phonon-magnon interaction, however, is not sufficient for a microscopic understanding of these parameter dependences of the SSE.

The importance of a microscopic understanding of the magnonic properties inside ferromagnetic insulators was also illustrated in a first series of magnetic-field dependence and length-scale probing measurements at ambient temperature \cite{Boona-2014,Kehlberger-2015,Kikkawa-2015,Ritzmann-2015,Jungfleisch-2015,Cornelissen-2016,Guo-2016}. Again, as the attached heat baths couple to the phonons inside the FMI only, the magnonic transport properties are exclusively driven by magnon-phonon coupling, so that it is crucial to study the magnon-phonon interaction processes inside the ferromagnet to understand the microscopic origin of the SSE transport properties.
Evidence of a ``phonon drag'' in the SSE was pointed out earlier in temperature-dependent measurements of the SSE, when the shape of the magnon conductivity showed the same temperature dependence as the corresponding phonon conductivity \cite{Adachi-2010,Jaworski-2011}. Recently, very direct evidence of the importance of the phonon-magnon interaction for the SSE was found in low-temperature measurements \cite{Kikkawa-2016} of the SSE in YIG, which showed sharp peaks in the spin Seebeck signal at the two specific magnetic fields where the magnon and phonon dispersions have touching points. This phenomenon was explained by the existence of ``magnon-polarons'' that describe mixed states which are neither purely magnonic nor phononic \cite{Flebus-2017}.

In this work we present a Boltzmann transport theory to describe the coupled magnon-phonon scattering mechanism in a simple model ferromagnetic insulator. In contrast to Ref. \onlinecite{Flebus-2017} we employ separate, incoherent magnon and phonon distributions, which, in the diffusive regime, may be described using an isotropic moment --- corresponding to a local phonon or magnon temperature --- and an anisotropic moment --- corresponding to a phonon or magnon (momentum) current density. We assume that the relaxation due to magnon-number non-conserving scattering processes, such as magnon dipole-dipole interaction, is faster than magnon-phonon scattering processes, so that no magnon chemical potential needs to be introduced \cite{Cornelissen-2016}. Despite the absence of coherence between magnon and phonon excitations, our findings qualitatively explain the experimental observation of peaks in the longitudinal spin Seebeck effect at low temperatures \cite{Kikkawa-2016}.

The remainder of this article is organized as follows: In Sec.\ \ref{sec:model} we present the Boltzmann equations for the magnons and phonons. We find microscopic expressions for the corresponding lifetimes from quantum-mechanically derived collision integrals. Upon imposing a hierarchy of ``fast'' and ``slow'' relaxation processes, the theory is reformulated in terms of a set of coupled hydrodynamic equations for the magnon and phonon distribution functions. The conversion of magnonic to electronic spin current at the ferromagnet--normal-metal interface is described using the spin-mixing conductance of the interface \cite{Tserkovnyak-2002}. In Sec.\ \ref{sec:results} we apply our theory to the ferrimagnetic insulator Y$_3$Fe$_5$O$_{12}$ (YIG), choosing model parameters such that the properties of a YIG$\vert$Pt heterostructure at low temperatures are well approximated \cite{Cherepanow-1993,Gurevich-1996}. We also present quantitative results for the relaxation rates, and transport coefficients of the magnon and phonon currents based on analytical evaluations, and compare our findings to the coherent magnon-polaron theory \cite{Flebus-2017}. In Sec.\ \ref{sec:conclusions} we present our conclusions.

\section{Model}
\label{sec:model}

\begin{figure}[t]
\includegraphics[width=\linewidth]{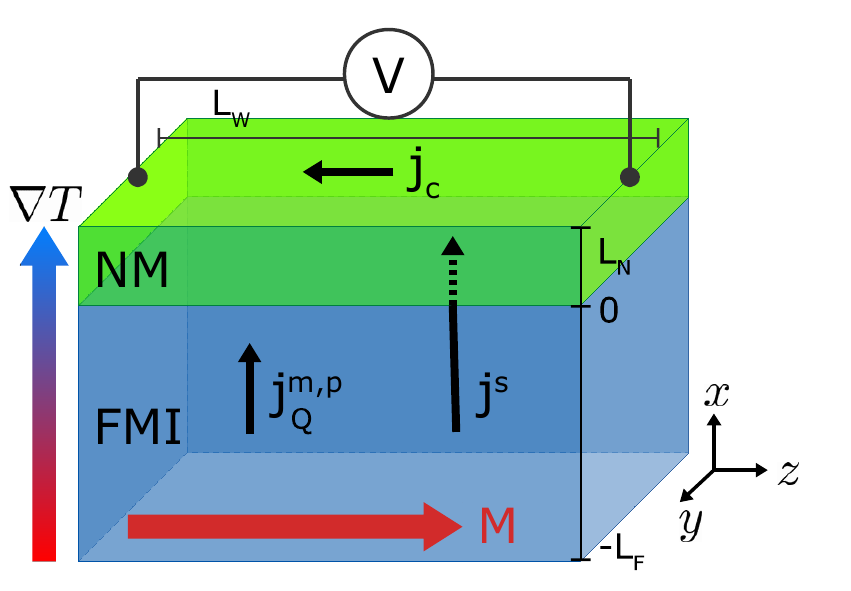}
\caption{\label{fig:setup} Illustration of the model setup for the longitudinal spin Seebeck effect. A ferromagnetic insulator of thickness $L_{\rm F}$ (bottom, blue) is coupled to a normal metal of thickness $L_{\rm N}$ (top, green). The coordinate axes are chosen such that the $x$ axis is perpendicular to the ferromagnet--normal metal interface plane. A temperature difference $\Delta T$ applied across the entire system results in the flow of a spin current $\vj^{\rm s}$ across the ferromagnet--normal metal interface, which can be measured by means of the inverse spin Hall effect.}
\end{figure}

We consider a system consisting of a ferromagnetic insulator of thickness $L_{\rm F}$ attached to a normal metal of thickness $L_{\rm N}$ as illustrated in Fig.\ \ref{fig:setup}. We choose coordinates such that the $x$ axis is perpendicular to the ferromagnet--normal metal interface, the ferromagnet and the normal metal occupying the space $-L_{\rm F} < x < 0$ and $0 < x < L_{\rm N}$, respectively. The system is coupled to heat baths on the left and the right, which are held at temperatures $T \pm \Delta T/2$, see Fig.\ \ref{fig:setup}. The magnetization direction and the applied magnetic field $\bf B$ are in the $z$-direction. We restrict ourselves to a low temperature regime, where umklapp scattering of magnons and phonons can be neglected, and optical magnons and phonons are frozen out. In YIG this corresponds to temperatures of a few K \cite{Sparks-1964,Gurevich-1996}.

The magnon and phonon distributions in the ferromagnetic insulator are described using their distribution functions $b_{\vk}(\vr,t)$ and $n_{\vq\lambda}(\vr,t)$, where $\vk$ and $\vq$ are the magnon and phonon wavevectors, respectively, and $\lambda$ denotes the phonon polarization. The distribution functions $b_{\vk}(\vr,t)$ and $n_{\vq\lambda}(\vr,t)$ satisfy coupled Boltzmann equations, which are the starting point of our analysis. 

\subsection{Magnon Boltzmann equation}

The Boltzmann equation for the magnon distribution $b_{\vk}(\vr,t)$ has the general form
\begin{align}
  \frac{\partial b_{\vk}(\vr,t)}{\partial t}
  + \vv_{\vk} \cdot \frac{\partial b_{\vk}(\vr,t)}{\partial \vr}
 = \left. \frac{db_{\vk}(\vr,t)}{dt}\right|_{\rm coll.}, \label{eq:mag_boltzmann}
\end{align}
where $\vv_{\vk}$ is the velocity of magnons with wavevector $\vk$. The collision term is separated into contributions from impurity/boundary scattering (i), magnon-magnon scattering (m), and magnon-phonon scattering (p),
\begin{align}
\left.\frac{db_{\vk}}{dt}\right|_{\rm coll.} = \left.\frac{d b_{\vk}}{dt}\right|_{\rm i} + \left.\frac{d b_{\vk}}{dt}\right|_{\rm m} + \left.\frac{d b_{\vk}}{dt}\right|_{\rm p}.  \label{eq:mag_collision}
\end{align}

For the impurity or boundary scattering contribution we use the relaxation-time form,
\begin{align}
\left.\frac{db_{\vk}(\vr,t)}{dt}\right|_{\rm i} = - \frac{b_{\vk}(\vr,t) - b^0_{\vk}(\vr,t)}{\tau^{\rm im}_{\vk}}. \label{eq:mag_impurity}
\end{align}
Here $b^0_{\vk}(\vr,t)$ is the equilibrium magnon distribution at temperature $T(\vr,t)$ and $\tau^{\rm im}_{\vk}$ is the relaxation time.

The magnon-magnon collision integral contains three-magnon processes which originate from dipole-dipole interaction as well as four-magnon processes which represent exchange scattering (see Tab.\ \ref{tab:diagrams}),
\begin{widetext}
\begin{align}
  \left.\frac{db_{\vk}}{dt}\right|_{\rm m} =&\, 
  \frac{2\pi}{\hbar} \sum_{\vk_2, \vk_1', \vk_2'} 
  |V^{\rm ex}(\vk_1', \vk_2';\vk , \vk_2)|^2 
  \delta(\varepsilon_{\vk} + \varepsilon_{\vk_2} - \varepsilon_{\vk_1'} - \varepsilon_{\vk_2'}) \delta_{\vk+\vk_2 - \vk_1' - \vk_2'}
  \nonumber \\ &\, \mbox{} \times
  \left[(1+b_{\vk})(1+b_{\vk_2})b_{\vk_1'} b_{\vk_2'} - b_{\vk} b_{\vk_2} (1+b_{\vk_1'})(1+b_{\vk_2'}) \right] \nonumber \\ &\, \mbox{}
  + \frac{2\pi}{\hbar} \sum_{\vk_2 , \vk'} |V^{\rm dip}(\vk , \vk_2 ; \vk')|^2 \delta(\varepsilon_{\vk} + \varepsilon_{\vk_2} - \varepsilon_{\vk'}) \delta_{\vk'-\vk-\vk_2}   \left[ (1+b_{\vk}) (1+b_{\vk_2}) b_{\vk'} - b_{\vk} b_{\vk_2} (1+b_{\vk'}) \right] \nonumber \\ &\, \mbox{}
  + \frac{\pi}{\hbar} \sum_{\vk_2' , \vk_1'} |V^{\rm dip}(\vk_2' , \vk_1' ;\vk)|^2 \delta(\varepsilon_{\vk} - \varepsilon_{\vk_2'} - \varepsilon_{\vk_1'}) \delta_{\vk-\vk_2'-\vk_1'}  \left[ (1+b_{\vk}) b_{\vk_2'} b_{\vk_1'} - b_{\vk} (1+b_{\vk_2'}) (1+b_{\vk_1'}) \right]. \label{eq:magnoncollision}
\end{align}
The first term on the right hand side of this expression represents the four-magnon processes, which are predominantly mediated by exchange processes. The corresponding symmetrized squared matrix element is \cite{Keffer-1961,Gurevich-1996}
\begin{align}
|V^{\rm ex}(\vk_1',\vk_2';\vk,\vk_2)|^2 = 2\left(\frac{g \mub D}{M V}\right)^2 ({\vk \cdot \vk_2})^2, \label{eqn:chapter2_magnon_magnon_exchange_potential}
\end{align}
where $D$ is the magnon exchange stiffness, $\mub$ the Bohr magneton, $g=2$ the Land\'e $g$-factor, $M$ the saturation magnetization, and $V$ the volume of the FMI. The second and third terms on the right hand side of Eq.\ (\ref{eq:magnoncollision}) account for magnon confluence processes and magnon splitting processes \cite{Kaganov-1961,Gurevich-1996}. The factor $1/2$ in the collision integral of the splitting processes was inserted to avoid double counting. These processes arise from dipole-dipole interactions or from relativistic effects and after symmetrization one has \cite{Gurevich-1996}
\begin{align}
  |V^{\rm dip}(\vk,\vk';\vk+\vk')|^2 =&\, \left(\frac{\mu_0}{4\pi}\right)^2 \frac{\pi^2 (g \mub)^3 M}{8V}
  \left| \frac{k_z k_{+}}{k^2} + \frac{k_z' k_{+}'}{k'^2}\right|^2, \label{eqn:chapter2_magnon_magnon_dipole_potential}
\end{align}
where $\mu_0$ is the vacuum permeability and $k_+ = k_x + i k_y$.

The third contribution to the magnon collision integral is from magnon-phonon collisions. The collision integral can be derived from the magneto-elastic Hamiltonian of Kaganov {\em et al.} \cite{Kaganov-1961,Kaganov-1961-2} and reads
\begin{align}
  \left.\frac{db_{\vk}}{dt}\right|_{\rm p} =&\,
  \frac{2\pi}{\hbar} \sum_{\vk',\vq,{\lambda}} |U(\vk,\vq,\lambda;\vk')|^2 \delta(\varepsilon_{\vk}+\omega_{\vq{\lambda}}-\varepsilon_{\vk'}) \delta_{\vk+\vq-\vk'} 
  \left[(1+b_{\vk})(1+n_{\vq{\lambda}}) b_{\vk'} - b_{\vk} n_{\vq{\lambda}} (1+b_{\vk'})\right] \nonumber\\ &\, \mbox{}
  + \frac{2\pi}{\hbar} \sum_{\vk',\vq',{\lambda}'} |U(\vk',\vq',\lambda';\vk)|^2 \delta(\varepsilon_{\vk}-\omega_{\vq'{\lambda'}} - \varepsilon_{\vk'}) \delta_{\vk-\vq'-\vk'} \left[(1+b_{\vk}) n_{\vq'{\lambda}'} b_{\vk'} - b_{\vk} (1+n_{\vq'{\lambda}'}) (1+b_{\vk'}) \right] \nonumber\\ &\, \mbox{}
  + \frac{2\pi}{\hbar} \sum_{\vk_2,\vq',{\lambda}'} |W^{(2)}(\vq',\lambda')|^2 \delta(\varepsilon_{\vk} - \omega_{\vq'{\lambda}'} + \varepsilon_{\vk_2}) \delta_{\vk+\vk_2-\vq'} \left[(1+b_{\vk})(1+b_{\vk_2}) n_{\vq'{\lambda}'} - b_{\vk} b_{\vk_2} (1+n_{\vq'{\lambda}'}) \right] \nonumber\\ &\, \mbox{}
  + \frac{2\pi}{\hbar} \sum_{\vq',{\lambda}'} |W^{(1)}(\vq',\lambda')|^2 \delta(\varepsilon_{\vk} - \omega_{\vq'{\lambda}'}) \delta_{\vk-\vq'} \left[ (1+b_{\vk}) n_{\vq'{\lambda}'} - b_{\vk} (1+n_{\vq'{\lambda}'}) \right],
  \label{eq:phononmagnon}
\end{align}
The squares $|U|^2$ and $|W|^2$ are expressed in terms of the magnon exchange stiffness $D$ and two magneto-elastic constants $B_{\parallel}$ and $B_{\perp}$ that represent dipole-dipole as well as spin-orbit interaction \cite{Kaganov-1961,Kaganov-1959}. The first two terms on the right hand side of this expression represent ``normal'' collision processes, in which the magnon number is conserved. The corresponding squared matrix element reads \cite{Rueckriegel-2014}
\begin{align}
  |U(\vk,\vq,\lambda;\vk')|^2 &= \frac{\hbar^2}{2 \varrho V \omega_{\vq_{\lambda}}} \left[ \frac{D}{2} \left((\vk\cdot \vq) \vk' + ({\vk'\cdot \vq) \vk} \right) \cdot \hat{e}_{\lambda} + \frac{B_{\parallel}}{S} (\vq - 3 q_z \hat{e}_z)\cdot \hat{e}_{\lambda} \right]^2, \label{eqn:chapter2_magnon_phonon_normal_amplitude}
\end{align}
where $\hat{e}_{\lambda}$ is the unit vector indicating the polarization direction of the phonon mode $(\vq,\lambda)$ and $S=M V_a / g \mub$ is the macrospin of a unit cell of volume $V_a$. The first term in this expression can also be derived from a Heisenberg model, by expanding the exchange couplings to lowest order in small displacements of the atomic positions, see Refs.\ \onlinecite{Kaganov-1958,Kaganov-1959}. 
The third term describes the pairwise creation or annihilation of magnons,
\begin{align}
  |W^{(2)}(\vq,{\lambda})|^2 &= \frac{\hbar^2}{2 \varrho V \omega_{\vq_{\lambda}}} \left[ \left(\frac{B_{\parallel}}{2S} (q_x \hat{e}_x - q_y \hat{e}_y) \cdot \hat{e}_{\lambda}\right)^2 + \left(\frac{B_{\perp}}{2S} (q_y \hat{e}_x + q_x \hat{e}_y) \cdot \hat{e}_{\lambda}\right)^2 \right]. \label{eqn:chapter2_magnon_phonon_relativistic_amplitude}
\end{align}
Finally, the fourth term on the right hand side of Eq.\ (\ref{eq:phononmagnon}) describes the conversion of a magnon into a single phonon and vice versa, with~\cite{Kaganov-1961-2,Rueckriegel-2014}
\begin{align}
  |W^{(1)}(\vq,{\lambda})|^2 &= \frac{\hbar^2}{2\varrho V_a \omega_{\vq_{\lambda}}} \frac{B^2_{\perp}}{2S} \left[ ((q_z \hat{e}_x + q_x \hat{e}_z)\cdot \hat{e}_{\lambda})^2 + ((q_z \hat{e}_y + q_y \hat{e}_z)\cdot \hat{e}_{\lambda})^2 \right]. \label{eqn:chapter2_magnon_phonon_conversion_amplitude}
\end{align}
In principle, the latter process gives rise to the existence of ``magnon polarons'' \cite{Kamra-2015,Shen-2015}, a coherent superposition of a magnon and phonon excitation. Sufficiently strong phonon-phonon and magnon-magnon scattering processes destroy the magnon-phonon coherence, however, validating our incoherent description in terms of the distribution function only. (Note that Ref.\ \onlinecite{Kikkawa-2016,Flebus-2017} uses a fully coherent description, finding results that do not differ qualitatively from ours.) The three types of magnon-phonon scattering processes are illustrated schematically in Table \ref{tab:diagrams}.

\begin{table}
\begin{ruledtabular}
\begin{tabular}{lcc}
process&
in&
out\\
\colrule
four-magnon (magnon interaction) & \includegraphics{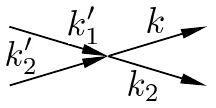} & \includegraphics{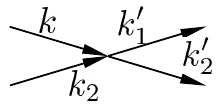} \\
three-magnon & \includegraphics{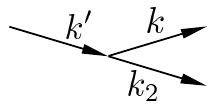} & \includegraphics{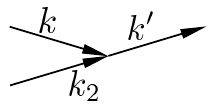} \\
 & \includegraphics{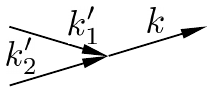} & \includegraphics{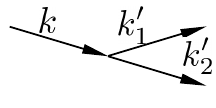} \\
\hline
phonon-to-magnon & \includegraphics{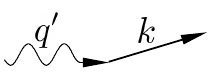} & \includegraphics{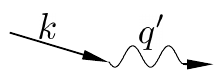} \\
phonon-to-two-magnon & \includegraphics{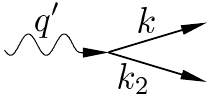} & \includegraphics{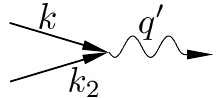} \\
magnon-phonon & \includegraphics{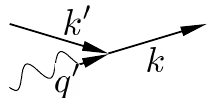} & \includegraphics{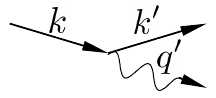} \\
 & \includegraphics{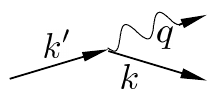} & \includegraphics{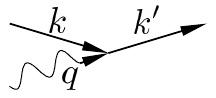} \\
\end{tabular}
\caption{\label{tab:diagrams} Schematic overview of the relevant magnon-magnon and magnon-phonon scattering processes.}
\end{ruledtabular}
\end{table}

\subsection{Phonon Boltzmann equation}

The Boltzmann equation for the phonon distribution function $n_{\vq\lambda}(\vr,t)$ reads
\begin{align}
  \frac{\partial n_{\vq\lambda}(\vr,t)}{\partial t}
  + \vc_{\vq \lambda} \cdot \frac{\partial n_{\vq\lambda}(\vr,t)}{\partial \vr}
 = \left. \frac{dn_{\vq\lambda}(\vr,t)}{dt}\right|_{\rm coll.}, \label{eq:ph_boltzmann}
\end{align}
where $\vc_{\vq\lambda}$ is the velocity of phonons with wavevector $\vq$ and polarization $\lambda$. The collision term is separated into contributions from impurity/boundary scattering (i), phonon-phonon scattering (p), and phonon-magnon scattering (m). As in the case of the magnons, we will describe phonon-impurity scattering using the relaxation-time approximation,
\begin{align}
\left. \frac{dn_{\vq\lambda}(\vr,t)}{dt} \right|_{\rm i} = - \frac{n_{\vq \lambda}(\vr,t) - n^0_{\vq\lambda}(\vr,t)}{\tau^{\rm ip}_{\vq{\lambda}}} \label{eq:ph_impurity}
\end{align}
where $\tau^{\rm ip}_{\vq{\lambda}}$ is the corresponding phonon-impurity scattering time. The expression for the phonon-magnon collision term reads
\begin{align}
  \left.\frac{dn_{\vq\lambda}}{dt}\right|_{\rm m} =&\,
  \frac{2\pi}{\hbar} \sum_{\vk,\vk'} |U(\vk,\vq,\lambda;\vk')|^2 \delta(\omega_{\vq{\lambda}}+ \varepsilon_{\vk}-\varepsilon_{\vk'}) \delta_{\vq+\vk-\vk'} 
  \left[(1+b_{\vk})(1+n_{\vq{\lambda}}) b_{\vk'} - b_{\vk} n_{\vq{\lambda}} (1+b_{\vk'})\right] \nonumber\\ &\, \mbox{}
  + \frac{\pi}{\hbar} \sum_{\vk_1',\vk_2'} |W^{(2)}(\vq,\lambda)|^2 \delta(\omega_{\vq{\lambda}} - \varepsilon_{\vk_1'} - \varepsilon_{\vk_2'}) \delta_{\vq-\vk_1'-\vk_2'} \left[b_{\vk_1'} b_{\vk_2'} (1+n_{\vq{\lambda}}) - (1+b_{\vk_1'})(1+b_{\vk_2'}) n_{\vq{\lambda}} \right] \nonumber\\ &\, \mbox{}
  + \frac{2\pi}{\hbar} \sum_{\vk'} |W^{(1)}(\vq,\lambda)|^2 \delta( \omega_{\vq{\lambda}}-\varepsilon_{\vk'}) \delta_{\vq-\vk'} \left[  b_{\vk'} (1+n_{\vq{\lambda}}) - (1+b_{\vk'}) n_{\vq{\lambda}} \right].
  \label{eq:phononmagnon_phonon}
\end{align}
\end{widetext}
We will not give an explicit expression for the phonon-phonon collision integral, since the corresponding collision rates do not enter in our final expressions (see the discussion below).

\subsection{Ansatz for the distribution function}
\label{sec:model_linearization}

To simplify the analysis of the coupled Boltzmann equations for the magnon and phonon distribution functions, we consider small deviations from equilibrium only and linearize the distribution functions $b_{\vk}(\vr,t)$ and $n_{\vq\lambda}(\vr,t)$ around the equilibrium distributions $b^0_{\vk} = 1/(e^{\varepsilon_{\vk}/k_B T}-1)$ and $n^0_{\vq\lambda} = 1/(e^{\omega_{\vq\lambda}/k_B T}-1)$,
\begin{align}
  b_{\vk}(\vr,t) =&\, b^0_{\vk} + \left( - \frac{\partial b^0_{\vk}}{\partial \varepsilon_{\vk}} \right) \varepsilon'_{\vk}(\vr,t), 
  \label{eq:ansatzb} \\
  n_{\vq\lambda}(\vr,t) =&\, n^0_{\vq\lambda} + \left( - \frac{\partial n^0_{\vq \lambda}}{\partial \omega_{\vq\lambda}} \right) \omega'_{\vq \lambda}(\vr,t).
\end{align}
We further assume that the magnon-magnon and phonon-phonon interactions are strong enough, when compared to the magnon-phonon interactions, that the magnon and phonon distributions everywhere are in a {\em local equilibrium}, characterized by energy and momentum densities $\rho_E^{\rm m,p}$ and $\rho_{\vk}^{\rm m,p}$. This corresponds to the parameterization \cite{Ziman-1960}
\begin{align}
  \varepsilon'_{\vk}(\vr,t) =&\, \frac{\varepsilon_{\vk}}{T} \Delta T^{\rm m} +  \vk \cdot \vv^{\rm m}, \nonumber \\
  \omega'_{\vq\lambda}(\vr,t) =&\, \frac{\omega_{\vq\lambda}}{T} \Delta T^{\rm p} + \vq \cdot \vv^{\rm p},
  \label{eq:ansatzp}
\end{align}
where $\Delta T^{\rm m,p}$ is the difference local magnon/phonon temperature and the (global) equilibrium temperature $T$ and $\vv^{\rm m,p}$ parameterizes the magnon/phonon momentum density. The temperature differences $\Delta T^{\rm m,p}$ and the velocities $\vv^{{\rm m,p}}$ are related to the corresponding energy and momentum densities as
\begin{align}
  \Delta \rho_{E}^{\rm m,p} &=\, C^{\rm m,p} \Delta T^{\rm m,p}, \nonumber \\
  \rho_{k_{\alpha}}^{\rm m,p} &=\, \sum_{\beta}
  {\cal C}^{\rm m,p}_{\alpha\beta} v^{\rm m,p}_{\beta},
\end{align}
which defines the specific heat capacities
\begin{align}
  C^{\rm m} &=\, \frac{1}{V} \sum_{\vk} \frac{\partial b^0_{\vk}}{\partial T} \, \varepsilon_{\vk}, \nonumber \\
  C^{\rm p} &=\, \frac{1}{V} \sum_{\vq,\lambda} \frac{\partial n^0_{\vq\lambda}}{\partial T} \, \omega_{\vq\lambda},
\end{align}
and the tensor coefficients
\begin{align}
  {\cal C}^{\rm m}_{\alpha\beta} &=\, \frac{1}{V} \sum_{\vk} \left( - \frac{\partial b^0_{\vk}}{\partial \varepsilon_{\vk}} \right) k_{\alpha} k_{\beta}, \nonumber \\
  {\cal C}^{\rm p}_{\alpha\beta} &=\, \frac{1}{V} \sum_{\vq,\lambda} \left( - \frac{\partial n^0_{\vq\lambda}}{\partial \omega_{\vq\lambda}} \right) q_{\alpha} q_{\beta}.
\end{align}
One verifies that the magnon-magnon and phonon-phonon collision integrals are zero for a distribution function of this form, since magnon-magnon and phonon-phonon collisions conserve energy and momentum. (Recall that we neglect umklapp processes.) 

The velocities $\vv^{\rm m,p}$ are related to the magnon spin current density $\vj^{\rm s}$ and heat current densities $\vj^{\rm m,p}_{Q}$,
\begin{align}
  \vj^{\rm s} =&\, \frac{\hbar}{V} \sum_{\vk} \vv_{\vk} b_{\vk}, \label{eq:jmagnon}  \\
  \vj^{\rm m}_{Q} =&\, \frac{1}{V} \sum_{\vk} \varepsilon_{\vk} \vv_{\vk} b_{\vk}, \label{eq:jqmagnon} \\
  \vj^{\rm p}_{Q} =&\, \frac{1}{V} \sum_{\vq,\lambda} \omega_{\vq\lambda} \vc_{\vq\lambda} n_{\vq\lambda}. \label{eq:jqphonon}
\end{align}
Upon substitution of the Ansatz (\ref{eq:ansatzb}) one finds
\begin{align}
  j^{\rm s}_{\alpha} =&\,
  \sum_{\beta} {\cal J}_{\alpha\beta} v^{\rm m}_{\beta} , \ \ j^{\rm m,p}_{Q\alpha} =\, \sum_{\beta} {\cal I}^{\rm m,p}_{\alpha\beta} v^{\rm m,p}_{\beta} 
\end{align}
with
\begin{align}
  {\cal J}_{\alpha\beta} =&\, \frac{\hbar}{V} \sum_{\vk} \left( - \frac{\partial b^0_{\vk}}{\partial \varepsilon_{\vk}} \right) v_{\vk\beta} k_{\alpha}, \nonumber \\
  {\cal I}^{\rm m}_{\alpha\beta} =&\, \frac{1}{V} \sum_{\vk} \left( - \frac{\partial b^0_{\vk}}{\partial \varepsilon_{\vk}} \right) \varepsilon_{\vk} v_{\vk\beta} k_{\alpha}, \nonumber \\
  {\cal I}^{\rm p}_{\alpha\beta} =&\, \frac{1}{V} \sum_{\vq,\lambda} \left( - \frac{\partial n^0_{\vq\lambda}}{\partial \omega_{\vq\lambda}} \right) \omega_{\vq\lambda} c_{\vq\lambda\beta} q_{\alpha}.
\end{align}
Impurity scattering and magnon-phonon scattering cause a further relaxation of the distribution functions. Impurity scattering tends to suppress any finite values of $\vv^{\rm m}$ and $\vv^{\rm p}$ but leaves $\Delta T^{\rm m}$ and $\Delta T^{\rm p}$ unaffected; magnon-phonon scattering suppresses differences $\Delta T^{\rm m} - \Delta T^{\rm p}$ and $\vv^{\rm m} - \vv^{\rm p}$. To derive the continuity equations for $\Delta T^{\rm m,p}$ and $\vv^{\rm m,p}$ we substitute the Ansatz (\ref{eq:ansatzp}) into the Boltzmann equations for the magnon and phonon distribution functions and calculate the rate of change of the energy and momentum densities. This gives
\begin{widetext}
\begin{align}
  \label{eq:evol1}
  C^{\rm m} \frac{\partial \Delta T^{\rm m}}{\partial t} + \sum_{\alpha,\beta} {\cal I}^{\rm m}_{\alpha\beta} \frac{\partial v^{\rm m}_{\alpha}}{\partial x_{\beta}} &=\, -G (\Delta T^{\rm m} - \Delta T^{\rm p}),  \\
  \label{eq:evol2}
  C^{\rm p} \frac{\partial \Delta T^{\rm p}}{\partial t} + \sum_{\alpha,\beta} {\cal I}^{\rm p}_{\alpha\beta} \frac{\partial v^{\rm p}_{\alpha}}{\partial x_{\beta}} &=\, -G (\Delta T^{\rm p} - \Delta T^{\rm m}),  \\
  \label{eq:evol3}
   \sum_{\alpha}  \left( \frac{{\cal I}^{\rm m}_{\beta\alpha}}{T} \frac{\partial \Delta T^{\rm m}}{\partial x_{\alpha}} + {\cal C}^{\rm m}_{\alpha\beta} \frac{\partial v^{\rm m}_{\alpha}}{\partial t} \right) &=\, - \sum_{\alpha} \left[{\cal G}^{\rm im}_{\alpha\beta} v^{\rm m}_{\alpha} + {\cal G}_{\alpha \beta} (v^{\rm m}_{\alpha} - v^{\rm p}_{\alpha}) \right],  \\
  \label{eq:evol4}
    \sum_{\alpha}  \left( \frac{{\cal I}^{\rm p}_{\beta\alpha}}{T} \frac{\partial \Delta T^{\rm p}}{\partial x_{\alpha}} + {\cal C}^{\rm p}_{\alpha\beta} \frac{\partial v^{\rm p}_{\alpha}}{\partial t} \right) &=\, - \sum_{\alpha} \left[{\cal G}^{\rm ip}_{\alpha\beta} v^{\rm p}_{\alpha} + {\cal G}_{\alpha \beta} (v^{\rm p}_{\alpha} - v^{\rm m}_{\alpha}) \right],  
\end{align}
where the tensor coefficients ${\cal I}^{\rm m,p}_{\alpha\beta} = \partial j_{Q\alpha}^{\rm m,p}/\partial v^{\rm m,p}_{\alpha} = \partial j^{\rm m,p}_{k_{\alpha}\beta}/\partial \Delta T^{\rm m,p}$, with $\vj_Q^{\rm m,p}$ and $\vj_{\vk}$ the energy and momentum current densities, respectively. The right hand side of Eqs.\ (\ref{eq:evol1})--(\ref{eq:evol4}) describes energy and momentum exchange between magnons and phonons and momentum exchange between magnons or phonons and impurities. The rates for these processes are
\begin{align}
  {\cal G}^{\rm im}_{\alpha\beta} &=\, \frac{1}{V} \sum_{\vk} \left( - \frac{\partial b^0_{\vk}}{\partial \varepsilon_{\vk}} \right) \frac{k_{\alpha} k_{\beta}}{\tau^{\rm im}_{\vk}},\ \ {\cal G}^{\rm ip}_{\alpha\beta} =\, \frac{1}{V} \sum_{\vq,\lambda} \left( - \frac{\partial n^0_{\vq\lambda}}{\partial \omega_{\vq\lambda}} \right) \frac{q_{\alpha} q_{\beta}}{\tau^{\rm ip}_{\vq\lambda}}, \nonumber \\
  G &=\, \frac{1}{V} \sum_{\vk,\vq,\lambda} \frac{\omega_{\vq\lambda}^2}{k_B T^2} \gamma_{\vk\vq\lambda},\ \
  {\cal G}_{\alpha\beta} =\, \frac{1}{V} \sum_{\vk,\vq,\lambda} \frac{q_{\alpha} q_{\beta}}{k_B T} \gamma_{\vk\vq\lambda},
\end{align}
where we abbreviated
\begin{align}
  \gamma_{\vk\vq\lambda} &=\,
  \frac{2 \pi}{\hbar} n^0_{\vq\lambda} \left[
  |U(\vk,\vq,\lambda;\vk+\vq)|^2 b^0_{\vk} (1 + b^0_{\vk+\vq}) 
    \delta(\omega_{\vq\lambda} + \varepsilon_{\vk} - \varepsilon_{\vk+\vq})
  \right. \nonumber \\ &\, \left. \mbox{}
  + \frac{1}{2}
    |W^{(2)}(\vq,\lambda)|^2 (1 + b^0_{\vq-\vk})(1+b^0_{\vk}) 
    \delta(\omega_{\vq\lambda} - \varepsilon_{\vk} - \varepsilon_{\vq-\vk})
  + |W^{(1)}(\vq,\lambda)|^2 (1+b^0_{\vk}) \delta_{\vq-\vk}
    \delta(\omega_{\vq\lambda} - \varepsilon_{\vk})
  \right].
\end{align}
\end{widetext}
Equations (\ref{eq:evol1})--(\ref{eq:evol4}) fully describe the coupled energy and momentum transport of the magnon and phonon systems. In the steady state, Eqs.\ (\ref{eq:evol3}) and (\ref{eq:evol4}) describe ``phonon drag'' and ``magnon drag'', the appearance of an anisotropic component of the magnon and phonon distributions in response to a gradient of the temperatures. The isotropic moment in Eqs.~(\ref{eq:evol1}) and (\ref{eq:evol2}) describes the relaxation of the magnon temperature towards the phonon temperature. 

\subsection{Boundary conditions and spin Seebeck voltage}

At the ferromagnetic insulator---insulator boundary at $x=-L_{\rm F}$ the magnon spin current (\ref{eq:jmagnon}) is zero, which implies $v^{\rm m}(-L_{\rm F}) = 0$, whereas the phonon temperature $\Delta T^{\rm p}(-L_{\rm F}) = \Delta T/2$ is determined by the temperature of the left heat bath, see Fig.\ \ref{fig:setup}. Similarly, at the normal-metal interface $x=0$ the phonon temperature $T^{\rm p}$ satisfies the boundary condition $\Delta T^{\rm p}(0) = -\Delta T/2$. The boundary condition for the magnon current at the normal-metal interface $x=0$ takes the form
\begin{align}
  j^{\rm s}_{x}(0) = S'_{\rm m} (\Delta T^{\rm m}(0) - \Delta T^{\rm e}(0)),
  \label{eq:jmboundary}
\end{align}
where $\Delta T^{\rm e}(0) = - \Delta T/2$ is the (deviation of the) electron temperature at the interface at $x=0$ and $S'_{\rm m}$ is the interface spin Seebeck coefficient \cite{Zhang-2012}, which can be expressed in terms of the real part $g_{\rm r}$ of the spin mixing conductance \cite{Xiao-2010},
\begin{align}
  S'_{\rm m} = \frac{g_{\rm r}}{\pi A S}
  \frac{1}{N} \sum_{\vk} \varepsilon_{\vk} \frac{d b_{\vk}^0}{dT},
  \label{eq:Sinterface}
\end{align}
where $S$ is the spin of the unit cell.
A derivation of the boundary condition (\ref{eq:jmboundary}) and a microscopic model leading to the expression for (\ref{eq:Sinterface}) for the interface spin Seebeck coefficient $S'_{\rm m}$ are given in appendix \ref{sec:interface}.

The relation between the spin current $j^{\rm s}_{x}(0)$ at the ferromagnet--normal metal interface and the transverse spin Seebeck voltage follows the theory of the inverse spin Hall effect. A nonzero spin current density implies a finite gradient of the spin accumulation $\mu_{\rm s} = \mu_{\uparrow} - \mu_{\downarrow}$ \cite{Zhang-2010-2}
\begin{align}
  \vj^{\rm s} = - \sigma_{\rm s} \partial_{\vr} \mu_{\rm s}, \label{eqn:chapter4_electron_spin_current_definition}
\end{align}
where $\sigma_{\rm s}$ is the spin conductivity. The spin accumulation satisfies the equation \cite{Zhang-2010-2}
\begin{align}
  \lambda_{\rm sf}^2 \partial_{\vr}^2 \mu_{\rm s}^2 = \mu_{\rm s}, \label{eqn:chapter4_spin_chemical_potential}
\end{align}
where $\lambda_{\rm sf}$ is the spin-flip length. Together with the boundary condition $j^{\rm s}_{x}(L_{\rm N}) = 0$ at the interface between the normal metal and the right heat bath, this gives the solution
\begin{align}
  j^{\rm s}_{x}(x) = j^{\rm s}_{x}(0) \frac{\sinh((L_{\rm N} - x)/\lambda_{\rm sf})}{\sinh(L_{\rm N}/\lambda_{\rm sf})}. \label{eqn:chapter4_electron_spin_current_solution}
\end{align}
The spin Seebeck voltage equals \cite{Xiao-2010} 
\begin{align}
  V_{\rm SSE}(x) = \frac{2 e}{\hbar} \theta_{\rm SH} L_{\rm W} \rho j^{\rm s}_{x}(x),
\end{align}
where $L_{\rm W}$ is the sample width, $\rho$ the electric resistivity, and $\theta_{\rm SH}$ is the spin Hall angle of the normal metal. Averaging over $x$ gives
\begin{align}
  V_{\rm SSE} =&\, \frac{1}{L_{\rm N}} \int_0^{L_{\rm N}} dx V_{\rm SSE}(x) \nonumber \\ =&\,
  \theta_{\rm SH} \rho \frac{2e}{\hbar} \frac{L_{\rm W}}{L_{\rm N}} \lambda_{\rm sf} j^{\rm s}_{x}(0) \tanh\left(\frac{L_{\rm N}}{2\lambda_{\rm sf}}\right). \label{eqn:chapter4_spin_seebeck_voltage}
\end{align}

\section{Results} 
\label{sec:results}

We use our theory to describe the longitudinal spin Seebeck effect in YIG$\vert$Pt heterostructures, where we put our focus on the magnetic field dependence measurements at low temperatures as performed by Kikkawa {\it et al.} \cite{Kikkawa-2015,Kikkawa-2016}. Since the longitudinal spin Seebeck effect is a steady-state phenomenon, driven by a time-independent temperature difference $\Delta T$ applied across the ferromagnet--normal-metal bilayer, we may neglect the time derivatives in the continuity equations (\ref{eq:evol1})--(\ref{eq:evol4}) and restrict our attention to time-independent solutions. Also, for the one-dimensional geometry of Fig.\ \ref{fig:setup}, all spatial dependences will be as a function of the coordinate $x$ only.

At low temperatures we may take a parabolic band for the magnon dispersion,
\begin{align}
\varepsilon_{\vk} = D |\vk|^2 + g \mub (B+ \mu_0 M), \label{eq:epsilon_k_ansatz}
\end{align}
with an offset due to the intrinsic exchange splitting and the Zeeman shift. Here $B$ denotes the applied magnetic field and $\mu_0 M$ is the exchange gap. For the phonons we restrict ourselves to the acoustic branches, 
\begin{align}
  \omega_{\vq\lambda} = \hbar c_{\lambda} |\vq|, \label{eq:omega_q_ansatz}
\end{align}
where $c_{\lambda}$ is the sound velocity and $\vq$ the phonon wave vector. For a YIG crystal oriented along the $\left\langle 100\right\rangle$ axis there are one longitudinal as well as two transverse polarized acoustic phonon branches. The values of the corresponding material properties which are used for our numerical calculation are summarized in Table \ref{tab:values}. 
For this simple model description the system is isotropic, so that the tensors ${\cal I}^{\rm m,p}$, ${\cal G}^{\rm im,ip}$, and ${\cal G}$ are proportional to the diagonal tensor.

Substituting the explicit expressions we find, for temperatures low enough that the dispersions (\ref{eq:epsilon_k_ansatz}) and (\ref{eq:omega_q_ansatz}) are valid for all thermally excited magnons and phonons, that
\begin{align}
  {\cal I}^{\rm m} = \frac{5}{16\hbar} \frac{(k_{\rm B} T)^{5/2}}{(\pi D)^{3/2}} e^{-\varepsilon_0 / (k_{\rm B} T)},
  \label{eq:Im}
\end{align}
with $\varepsilon_0 = g \mub (B + \mu_0 M)$,
and
\begin{align}
  {\cal I}^{\rm p} = \sum_{\lambda} \frac{2 \pi^2 (k_{\rm B} T)^4}{45 c_{\lambda}^3 \hbar^4}.
  \label{eq:Ip}
\end{align}

\begin{table}
\begin{ruledtabular}
\begin{tabular}{lcr}
\textrm{Quantity}&
\textrm{Value}&
\textrm{Reference}\\
\colrule
$a$ (YIG) & 1.24 nm & \cite{Gurevich-1996}\\
$D$ & 8.5$\times$10$^{-40}$ J\,m$^2$ & \cite{Shinozaki-1961}\\
$\mu_0 M_{\rm s}$ & 0.18 T & \cite{Kajiwara-2010}\\
$c_{\parallel}$ & 7209 m/s & \cite{Rueckriegel-2014} \\
$c_{\perp}$ & 3843 m/s & \cite{Rueckriegel-2014} \\
$\varrho$ & 21450 kg/m$^3$  & \cite{Cherepanow-1993} \\
$B_{\parallel}$ & 4.12$\times$10$^{-3}$ eV & \cite{Rueckriegel-2014} \\
$B_{\perp}$ & 8.24$\times$10$^{-3}$ eV & \cite{Rueckriegel-2014} \\
\hline
$a$ (Pt) & 0.39 nm & \cite{Ashcroft-1976} \\
$g_r /A$ & 10$^{16}$ 1/m$^2$ & \cite{Kajiwara-2010} \\
$\theta_{\rm SH}$ & 0.0037 & \cite{Maekawa-2007} \\
$\lambda_{\rm sf}$ & 7.3 nm & \cite{Du-2015} \\
$\rho$ & 0.91$\times$10$^{-6}$ $\Omega$/m & \cite{Uchida-2010-2} \\
$L_{\rm N} \times L_{\rm W}$ & 5 nm $\times$ 2 mm & \\
\end{tabular}
\caption{\label{tab:values} Parameter values used for the numerical calculations. The third column lists the relevant references where these values were obtained.}
\end{ruledtabular}
\end{table}

In the steady-state limit the velocities $\vv^{\rm m}$ and $\vv^{\rm p}$ can be obtained from Eqs.~(\ref{eq:evol3}) and (\ref{eq:evol4}),
\begin{align}
  \vv^{\rm m} =&\, - \frac{1}{T} \frac{({\cal G}^{\rm ip} + {\cal G}) {\cal I}^{\rm m} \partial_{\vr} \Delta T^{\rm m} + {\cal G} {\cal I}^{\rm p} \partial_{\vr} \Delta T^{\rm p}}{{\cal G}^{\rm im} {\cal G^{\rm ip} + {\cal G} {\cal G}^{\rm im} + {\cal G} {\cal G}^{\rm ip}}}, \label{eq:vmgradient} \\
  \vv^{\rm p} =&\, - \frac{1}{T} \frac{({\cal G}^{\rm im} + {\cal G}) {\cal I}^{\rm p} \partial_{\vr} \Delta T^{\rm p} + {\cal G} {\cal I}^{\rm m} \partial_{\vr} \Delta T^{\rm m}}
  {{\cal G}^{\rm im} {\cal G^{\rm ip} + {\cal G} {\cal G}^{\rm im} + {\cal G} {\cal G}^{\rm ip}}},
  \label{eq:vpgradient}
\end{align}
They imply a relation between the magnon spin current $\vj^{\rm s}$, and between the magnon and phonon heat currents $\vj^{\rm m,p}_Q$ and the gradients of the magnon and phonon temperatures,
\begin{align}
  \vj^{\rm s} =&\, -S_{\rm m} \partial_{\vr} \Delta T^{\rm m} - S_{\rm d} \partial_{\vr} \Delta T^{\rm p} , \\
  \vj^{\rm m,p}_Q =&\, - \kappa_{\rm m,p} \partial_{\vr} \Delta T^{\rm m,p} - \kappa_{\rm d} \partial_{\vr} \Delta T^{\rm p,m} ,
\end{align}
with the (bulk) spin Seebeck coefficients
\begin{align}
  S_{\rm m} =&\, \frac{1}{T} \frac{{\cal J} ({\cal G}^{\rm ip} + {\cal G}) {\cal I}^{\rm m}} {{\cal G}^{\rm im} {\cal G^{\rm ip} + {\cal G} {\cal G}^{\rm im} + {\cal G} {\cal G}^{\rm ip}}}, \nonumber \\ 
  S_{\rm d} =&\, \frac{1}{T} \frac{{\cal J} {\cal G} {\cal I}^{\rm p}}
  {{\cal G}^{\rm im} {\cal G^{\rm ip} + {\cal G} {\cal G}^{\rm im} + {\cal G} {\cal G}^{\rm ip}}},
\end{align}
and thermal conductivities
\begin{align}
\kappa_{\rm m} =&\, \frac{1}{T} \frac{({\cal G}^{\rm ip} + {\cal G}) ({\cal I}^{\rm m})^2} {{\cal G}^{\rm im} {\cal G^{\rm ip} + {\cal G} {\cal G}^{\rm im} + {\cal G} {\cal G}^{\rm ip}}}, \nonumber \\ 
\kappa_{\rm p} =&\, \frac{1}{T} \frac{({\cal G}^{\rm im} + {\cal G}) ({\cal I}^{\rm p})^2} {{\cal G}^{\rm im} {\cal G^{\rm ip} + {\cal G} {\cal G}^{\rm im} + {\cal G} {\cal G}^{\rm ip}}}, \nonumber \\ 
\kappa_{\rm d} =&\, \frac{1}{T} \frac{{\cal G} {\cal I}^{\rm p} {\cal I}^{\rm m}}
  {{\cal G}^{\rm im} {\cal G^{\rm ip} + {\cal G} {\cal G}^{\rm im} + {\cal G} {\cal G}^{\rm ip}}} .
\end{align}
The coefficients $S_{\rm d}$ and $\kappa_{\rm d}$ describe the ``phonon drag'', the anisotropic component of the magnon distribution in response to a gradient of the phonon temperature.

\begin{figure}
\includegraphics[width=\linewidth]{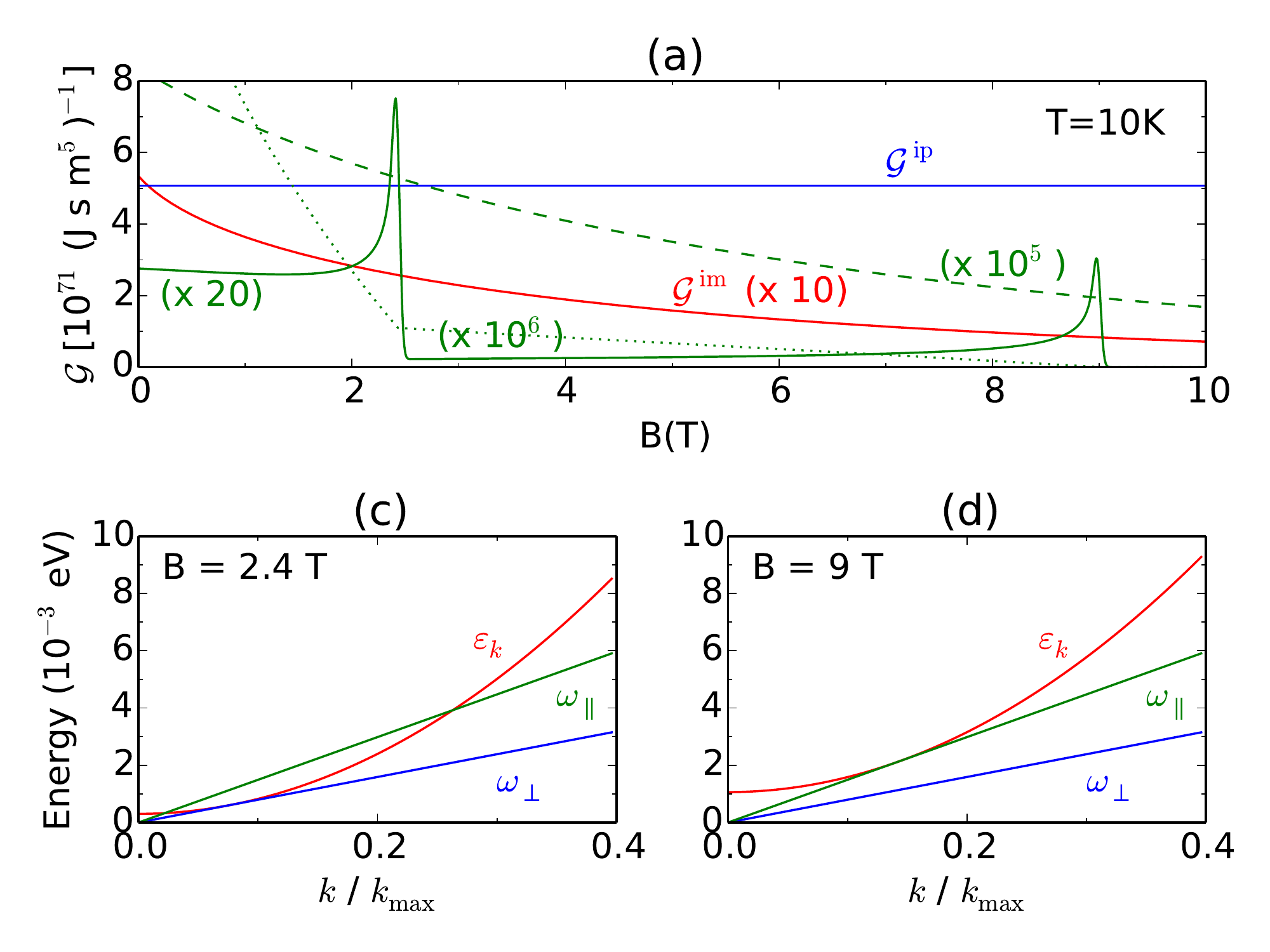}
\caption{\label{fig:2} (a) Three contributions to the magnon-phonon relaxation rate ${\cal G}$, the magnon-impurity rate ${\cal G}^{\rm im}$, and the phonon-impurity rate ${\cal G}^{\rm im}$, as a function of the applied magnetic field $B$. The three contributions to the magnon-phonon rate are from magnon-phonon-scattering (solid), phonon-to-magnon conversion (dashed), and phonon-to-two-magnon conversion (dotted). The magnon and phonon energy dispersions for the ``critical'' applied magnetic fields $B_{\perp} = 2.4\;{\rm T}$ and $B_{\parallel} = 9\;{\rm T}$ are shown in panels (b) and (c), respectively.}
\end{figure}

To obtain a numerical estimate of the relevant relaxation rates, we use the values given in Table \ref{tab:values}. For impurity scattering we assume that the phonon-impurity and magnon-impurity scattering times $\tau^{\rm ip}_{\vq\lambda} = \tau_{\rm ip}$ and $\tau^{\rm im}_{\vk} = \tau_{\rm im}$ are independent of $\vq$ and $\lambda$ and $\vk$, respectively, and extract these times from the low-temperature thermal conductivity and its magnetic field dependence as in Boona {\it et al.}\ \cite{Boona-2014}. This gives the values $\tau_{\rm im} = 3.4 \times 10^{-8}\;{\rm s}$ and $\tau_{\rm ip} = 5.6\times 10^{-8}\;{\rm s}$. Figure \ref{fig:2} shows the temperature and magnetic-field dependences of the three contributions to the magnon-phonon relaxation rates ${\cal G}$ and of the impurity rates ${\cal G}^{\rm im}$ and ${\cal G}^{\rm ip}$.

The phonon-to-magnon conversion and phonon-to-two-magnons conversion cease to contribute to the relaxation rate ${\cal G}$ above a ``critical'' magnetic field, because of energy and momentum conservation considerations. Since the phonon and magnon dispersions are tangential at the critical magnetic field, see Fig.\ \ref{fig:2}c and d, the magnon-to-phonon conversion rates diverges upon approaching the critical field from below. There are two such divergences, because longitudinal and transversal phonons have different velocities and, hence, different critical fields. The divergences should be broadened by the finite phonon-phonon and magnon-magnon scattering times, which are not considered explicitly in the continuity equations (\ref{eq:evol1})--(\ref{eq:evol4}). (Instead, phonon-phonon and magnon-magnon scattering are considered implicitly, because they enforce the local equilibrium form (\ref{eq:ansatzp}) of the magnon and phonon distribution functions.) Alternatively, the divergence is partially suppressed once deviations from the fully isotropic dispersions (\ref{eq:epsilon_k_ansatz}) and (\ref{eq:omega_q_ansatz}) are taken into account. We broaden the divergence by an average magnon-magnon lifetime $\tau_{\rm m} = 10^{-10}\;{\rm s}$ \ \cite{Rezende-2014} due to magnon number conserving scattering which corresponds to an energy broadening $\epsilon = \hbar / 2 \tau_{\rm m} = 10^{-6}\;{\rm eV}$. The (relativistic) phonon-to-two-magnons conversion shows a monotonous decay upon increasing the magnetic field.
  
The isotropic moment in Eqs.\ (\ref{eq:evol1}) and (\ref{eq:evol2}) describes the relaxation of the magnon temperature towards the phonon temperature. Of special interest for the calculation of the spin Seebeck current is the so-called thermal decay length $\lambda$, which describes the length scale over which a difference between magnon and phonon temperatures at the interface relaxes towards the center of the ferromagnet. Upon substituting the velocities (\ref{eq:vmgradient}) and (\ref{eq:vpgradient}) into the continuity equations (\ref{eq:evol1}) and (\ref{eq:evol2}) we obtain two second order differential equation for the magnon and phonon temperatures,
\begin{align}
&\kappa_{\rm m} \partial^2_x \Delta T^{\rm m} + \kappa_{\rm d} \partial^2_x \Delta T^{\rm p} = - \kappa_{\rm p} \partial^2_x \Delta T^{\rm p} - \kappa_{\rm d} \partial^2_x \Delta T^{\rm m} \nonumber\\
&= G (\Delta T^{\rm m} - \Delta T^{\rm p}) .
\end{align}  
The general solution of these equations is of the form
\begin{align}
  \Delta T^{\rm m}(x) =&\, \Delta T_0 + \alpha x + (\kappa_{\rm p} + \kappa_{\rm d}) \sum_{\pm} \delta_{\pm} e^{\pm x/\lambda} \nonumber \\
  \Delta T^{\rm p}(x) =&\, \Delta T_0 + \alpha x - (\kappa_{\rm m} + \kappa_{\rm d}) \sum_{\pm} \delta_{\pm} e^{\pm x/\lambda} ,
  \label{eq:DTgeneral}
\end{align}
with the decay length
\begin{align}
\lambda^2 = \frac{\kappa_{\rm m} \kappa_{\rm p} - \kappa^2_{\rm d}}{G (\kappa_{\rm p} + \kappa_{\rm m} + 2 \kappa_{\rm d})}
\end{align}
and with coefficients $\Delta T_0$, $\alpha$, and $\delta_{\pm}$ that are determined by the boundary conditions.

Taking the parameter values for YIG, see Table \ref{tab:values}, we find that find that $\kappa_{\rm p} \gg \kappa_{\rm m,d}$. In the limit $\kappa_{\rm p} \gg \kappa_{\rm m,d}$, Eq.\ (\ref{eq:DTgeneral}) gives a strictly linear spatial profile for the phonon temperature, so that $\Delta T_0 = 0$ and $\alpha = -\Delta T/L_{\rm F}$. The remaining parameters $\delta_{\pm}$ can then be obtained from the boundary conditions for the magnon spin current at $x=-L_{\rm F}$ and $x=0$. The result for the (magnon) spin current $j^{\rm s}_{x}(0)$ at the ferromagnet--normal metal interface is
\begin{align}
  j^{\rm s}_{x}(0) = \frac{\Delta T}{L} \frac{(S_{\rm m} + S_{\rm d}) \tanh(L/2 \lambda)}{S_{\rm m}/(\lambda S'_{\rm m}) + \coth(L/\lambda)}. \label{eq:jsresult}
\end{align}
Substitution of Eq.\ (\ref{eq:jsresult}) into the expression (\ref{eqn:chapter4_spin_seebeck_voltage}) gives the corresponding spin Seebeck voltage. An analytic solution of the equations is possible without the simplifications associated with the limit $\kappa_{\rm p} \gg \kappa_{\rm m,d}$, too, although the resulting expression for $j^{\rm s}_{x}(0)$ is not as concise as Eq.\ (\ref{eq:jsresult}). 

\begin{figure}
\includegraphics[width=\linewidth]{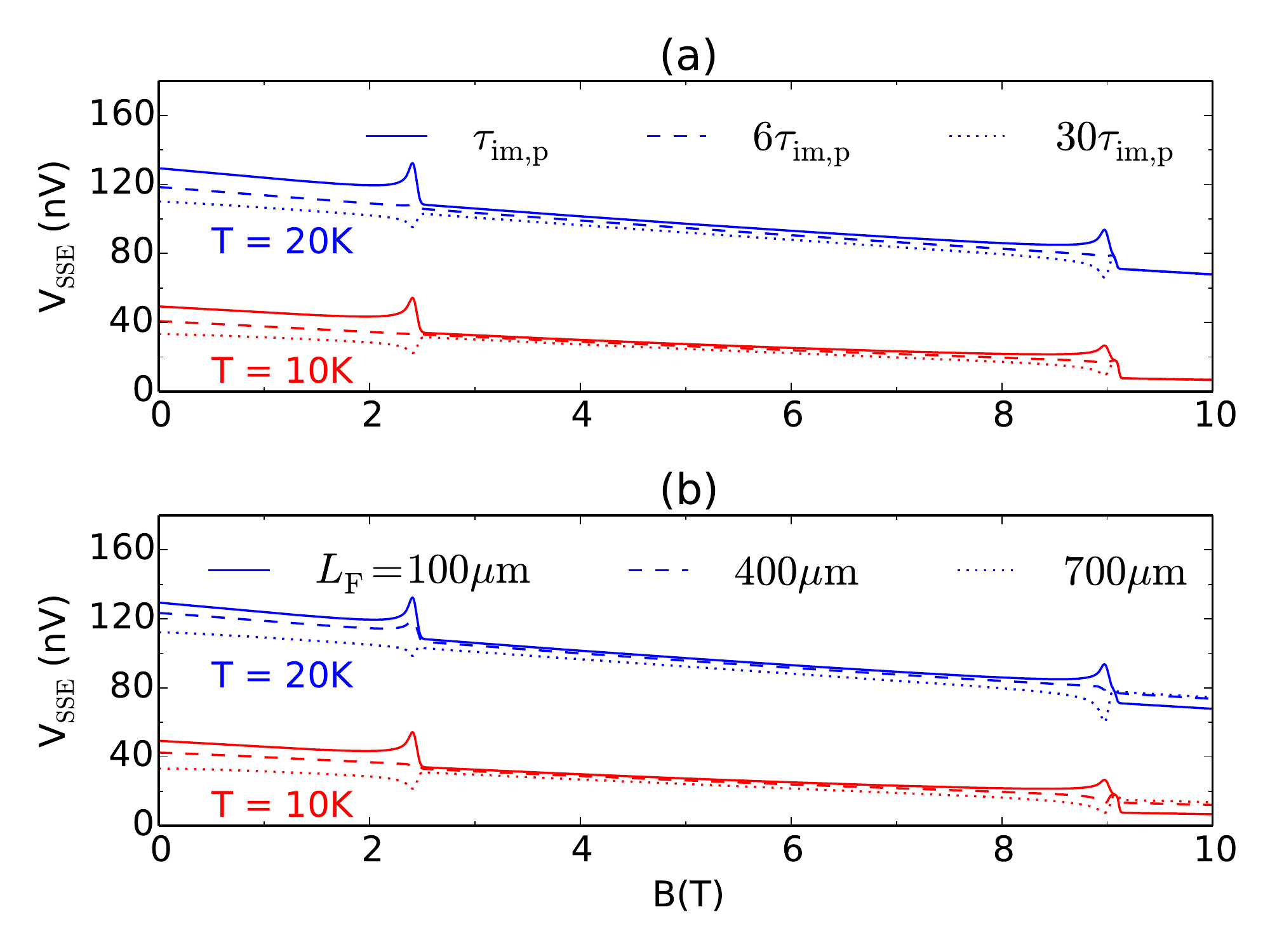}
\caption{\label{fig:3}Spin Seebeck voltage as a function of the applied magnetic field for different temperatures. (a) Three magnon/phonon impurity rates $\tau_{\rm im,p}$ at fixed length $L_{\rm F} = 100\mu{\rm m}$. (b) Three different lengths for fixed $\tau_{\rm im,p}$ as in Tab.\ \ref{tab:values}.}
\end{figure}

Figure \ref{fig:3} shows the resulting spin Seebeck voltage $V_{\rm SSE}$ for two different temperatures, as a function of the applied magnetic field. Although $V_{\rm SSE}$ generally decreases upon increasing the magnetic field, sharp features exist near the ``critical'' magnetic fields at which the magnon-to-phonon conversion rate diverges. The magnitude of those features depends on the impurity lifetimes $\tau_{\rm im,p}$, and the length of the FMI. These results are similar to those obtained by Kikkawa {\em et al.} and Flebus {\em et al.} \cite{Kikkawa-2016,Flebus-2017}, where the magnon-to-phonon conversion processes were treated coherently. The incoherent approach taken here should be valid if the incoherent processes dominate over the coherent ones, {\em i.e.}, if the magnon-to-phonon conversion matrix elements are small in comparison to the phonon-phonon and/or magnon-magnon scattering lifetimes.
At a fixed applied magnetic field the dependence of the longitudinal spin Seebeck effect on the thickness $L_{\rm F}$ of the ferromagnetic layer is governed by the combination $\lambda S'_{\rm m}/S_{\rm m}$, which controls the experimentally observed saturation of the LSSE signal towards bulk ferromagnetic samples \cite{Kehlberger-2015}.

\section{Conclusions}
\label{sec:conclusions}

In summary, we constructed a Boltzmann description for the coupled magnon-phonon transport in a simple model ferromagnetic insulator. In our description the magnon-phonon coupling is accounted for explicitly through its appearance in the collision integrals. Phonon-phonon and magnon-magnon relaxation processes, on the other hand, are taken into account implicitly, as they impose a local-equilibrium form of the magnon and phonon distribution functions. The magnon-phonon coupling leads to a ``phonon drag'' contribution to the magnon spin current in the ferromagnetic insulator. 

At low temperatures, of the three types of magnon-phonon coupling terms that we consider --- phonon-to-magnon conversion, phonon-to-two-magnon conversion, and magnon-phonon interaction --- the first process causes sharp peaks or dips in the magnon-phonon scattering rate at a critical magnetic field strength, where the magnon and phonon dispersions have touching points. In general there are two such critical magnetic field strengths, corresponding to longitudinal and transverse phonon branches. Whether a peak or dip is observed depends on the relative magnitudes of the other scattering rates involved, such as magnon-impurity and phonon-impurity rates, and the size of the ferromagnetic sample. These findings agree qualitatively with the experimental observations of Kikkawa {\em et al.}\ \cite{Kikkawa-2016}. In particular, our incoherent Boltzmann approach yields similar features for the resulting spin Seebeck voltage as the theory of Refs. \onlinecite{Kikkawa-2016,Flebus-2017}, which used a fully coherent coupling of magnon and phonon systems, leading to the formation of ``magnon-polarons''.

In particularly in the limit of thick ferromagnetic layers, the strength of the spin Seebeck effect may depend strongly on the properties of the ferromagnet--normal-metal interface. We expanded the spin-pumping interface model of Xiao {\it et al.}, which uses the spin-mixing conductance to characterize the interface properties, beyond the classical high temperature limit, to describe the low temperature regime. 
In addition we showed the consistency of the spin mixing conductance model to the alternatively used description of the interface coupling in terms of an $sd$-like exchange coupling between magnons in the ferromagnet and spin-polarized electrons in the normal metal as in Ref. \onlinecite{Zhang-2010}.

Possible extensions of our theory include a more microscopic treatment of the magnon-magnon interactions, relaxing the local-equilibrium assumptions made in our present description, or the inclusion of time-dependent effects. In particular, we can use our approach to investigate the temporal evolution of spatially inhomogeneous magnon and phonon temperatures and investigate the relevant time scales, that govern the evolution of the spin Seebeck effect on short time scales \cite{Seifert-2018}. We leave such extensions for future work.

\section*{Acknowledgement}

This paper was financially supported by the Deutsche Forschungsgesellschaft within the Priority Program SPP 1538 "Spin-caloric Transport" and the Collaborative Research Center TRR 227 ``Ultrafast Spintronics''. We would like to thank G.E.W. Bauer for stimulating discussions.


\appendix

\section{Magnon-electron coupling at FN interface}
\label{app:a}

As in the main text, we consider a ferromagnetic insulator occupying the region $-L_{\rm F} < x < 0$, coupled to a normal metal at $0 < x < L_{\rm N}$. If the normal metal is at zero temperature, the spin current density at the ferromagnetic insulator--metal interface is \cite{Tserkovnyak-2002}
\begin{equation}
  j^{\rm s}_x = \frac{\hbar}{4 \pi} \frac{g_{\rm r}}{A} \left\langle\vm \times \dot \vm\right\rangle_{x},
\end{equation}
where $g_{\rm r}$ is the real part of the spin-mixing conductance and $\vm$ is a unit vector pointing in the direction of the magnetization at the interface. (We omit a second contribution to $j^{\rm s}_x$, which is proportional to the imaginary part of the spin-mixing conductance and vanishes upon time averaging.) Expressing $\vm(\vr)$ in terms of magnon creation and annihilation operators $a^{\dagger}(\vr)$ and $a(\vr)$ for a macrospin of magnitude $S$ located at position $\vr$ and normal ordering, one finds
\begin{align}
\left\langle \vm(\vr) \times \dot \vm(\vr) \right\rangle_{x} &=\, \frac{i}{S} \langle \dot a(\vr)^{\dagger} a(\vr) - a(\vr)^{\dagger} \dot a(\vr) \rangle.
\end{align}
Substituting the mode expansion
\begin{equation}
  a(\vr) = \sum_{\vk} \sqrt{\frac{2-\delta_{\vk,0}}{N}} \cos(k_x x) e^{i k_y y + i k_z z} a_{\vk}, \label{eq:aexpand}
\end{equation}
where $N$ is the number of macrospins in the ferromagnetic insulator and the position $\vr$ is taken at the center of the unit cell, we find, for $\vr$ at the ferromagnetic insulator--normal metal interface at $x=0$,
\begin{equation}
 \left\langle \vm(\vr) \times \dot \vm(\vr) \right\rangle_{x} = \frac{2}{N S} \sum_{\vk} \frac{(2-\delta_{\vk,0})\varepsilon_{\vk}}{\hbar} b^0_{\vk},
\end{equation}
so that
\begin{equation}
  j^{\rm s}_x = \frac{g_{\rm r}}{\pi A S} \frac{1}{2N} \sum_{\vk} (2-\delta_{\vk,0}) \varepsilon_{\vk} b^0_{\vk}. \label{eq:js0}
\end{equation}

If the normal metal is at a finite temperature, the net spin current across the ferromagnet--normal metal interface is the sum of the spin current (\ref{eq:js0}) and a counterflow temp given by the same expression, but with the magnon temperature $T_{\rm m}$ replaced by the electron temperature $T_{\rm e}$. Hence, with a small temperature difference $\Delta T = T_{\rm m} - T_{\rm e}$ between normal metal and ferromagnetic insulator, one finds
\begin{equation}
  j^{\rm s}_x = - S_{\rm m}' \Delta T,
\end{equation}
with
\begin{equation}
  S_{\rm m}' = \frac{g_{\rm r}}{\pi A S} \frac{1}{2 N} \sum_{\vk} (2-\delta_{\vk,0}) \varepsilon_{\vk} \frac{db^0_{\vk}}{dT} .
  \label{eq:Spgeneral}
\end{equation}
This is the same expression as Eq.\ (\ref{eq:Sinterface}) of the main text, where the limit of a macroscopic sample with no exception for $\vk=0$ was taken.

\section{Simple model for magnon-electron coupling at FN interface}
\label{sec:interface}

\begin{figure}
\includegraphics{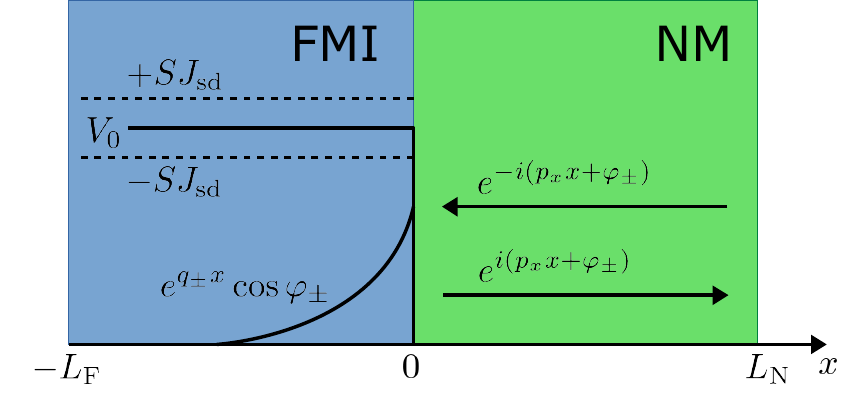}
\caption{Simple interface model: Electrons incident from the normal metal ($x > 0$) reflect from the ferromagnetic insulator with a scattering phase shift $\varphi_{\pm}$ for majority ($+$) and minority ($-$) electrons. The ferromagnetic insulator is modeled by a band offset $V_0$ larger than the Fermi energy, ensuring that the electron wavefunction decays exponentially inside the insulator.}
\label{fig:spinmixing}
\end{figure}

To make the general expression (\ref{eq:Spgeneral}) concrete, we describe the interface between a ferromagnetic insulator (for $x < 0$) and a normal metal (for $x > 0$) using the Hamiltonian
\begin{align}
  {\cal H} = \frac{\hbar^2 (\vp^2 - p_{\rm F}^2)}{2 m} + V(x) + J_{\rm sd} \vS(\vr) \cdot \vsigma,
\end{align}
where $\vp$ is the electron wavevector, $p_{\rm F}$ the Fermi wavenumber, $m$ the electron mass, $J_{\rm sd}$ the $sd$ exchange interaction between conduction electrons and localized spins, $V(x)$ a potential chosen such that $V(x) = V_0 > \hbar^2 p_{\rm F}^2/2m$ in the ferromagnetic insulator (for $x < 0$) and $V(x) = 0$ in the normal metal (for $x > 0$), see Fig.\ \ref{fig:spinmixing}. The spin operator 
\begin{equation}
  \vS(\vr) = \frac{1}{V_a} \sum_{j} \vS_j \delta(\vr-\vr_j),
\end{equation}
where $\vS_j$ is the macrospin operator and $V_a$ is the size of the unit cell. For small deviations from a uniform magnetization in the $z$ direction we may write
\begin{equation}
  S_{jz} = S,\ \
  S_{j+} = \sqrt{2 S} a(\vr_j),\ \
  S_{j-} = \sqrt{2 S} a(\vr_j)^{\dagger} ,
\end{equation}
where $\vr_j$ is the center of the $j$th unit cell, $S$ is the size of the macrospin, and $a(\vr)$ the magnon operator, which has the mode expansion given in Eq.\ (\ref{eq:aexpand}). For this model, we now calculate the mixing conductance $g_{\rm r}$ and the spin current $j^{\rm s}_x(0)$.

{\em Calculation of the mixing conductance.---}
The wavefunction of an electron with wavevector $\vp_{\parallel}$ to the interface, $|\vp| = p_{\rm F}$, and spin $\pm$ is
\begin{align}
  \psi_{\pm}(x) = \sqrt{\frac{2}{V_{\rm N}}} e^{i \vp_{\parallel} \cdot \vr} 
  \times \left\{ \begin{array}{ll}
    \cos(p_x x + \varphi_{\pm}) & \mbox{if $x > 0$}, \\
   e^{q_{\pm} x} \cos \varphi_{\pm}
  & \mbox{if $x < 0$}, \end{array} \right.
\end{align}
where $V_{\rm N}$ is the volume of the normal metal, $q_{\pm}^2 = 2 m (V_0 \pm S J_{\rm sd}) - p_x^2$, $\tan \varphi_{\pm} = -q_{\pm}/p_x$, and $p_x^2 = p_{\rm F}^2 - p_{\parallel}^2$. Replacing $\vS(\vr)$ by $S \ve_z$, we find for the real part of the mixing conductance
\begin{align}
  g_{\rm r} =&\, 2 \sum_{\vp_{\parallel}} \sin^2(\varphi_{+}-\varphi_{-})
  \nonumber \\
  \approx &\, 2 \sum_{\vp_{\parallel}} \frac{(S J_{\rm sd})^2 p_x^2}{V_0^2 (2 m V_0 - p_x^2)},
  \label{eq:gr}
\end{align}
where in the second equality we expanded to second order in $J_{\rm sd}$. 

{\em Calculation of the spin current $j^{\rm s}_x(0)$.---} We calculate the spin current to order $J_{sd}^2$ using the Fermi Golden rule. From the matrix element
\begin{align}
  \langle \psi'_-,n_{\vk}-1| {\cal H} | \psi_+,n_{\vk} \rangle 
  =&\, \frac{p_x^2 J_{\rm sd} \sqrt{S}}{m V_0 L_{\rm N} \sqrt{N(2 m V_0 - p_x^2)}}
  \nonumber \\ &\, \mbox{} \times
 \sqrt{n_{\vk}}\, \delta_{\vp_{\parallel}'=\vp_{\parallel}+\vk_{\parallel}},
  \label{eq:matrixelement}
\end{align}
where $|\psi_{+},n_{\vk}\rangle$ ($|\psi'_{-},n_{\vk}-1 \rangle$) is a state with an electron of spin $+$ ($-$) and transverse wavevector $\vp_{\parallel}$ ($\vp_{\parallel}'$) and $n_{\vk}$ ($n_{\vk}-1$) magnons in the model labeled by the wavevector $\vk$. In the calculation of the matrix element (\ref{eq:matrixelement}) we assumed that all magnon momenta are small in comparison to the Fermi momentum, so that we may neglect the change of the magnitude of the longitudinal wavevector component $p_x$ of the electrons upon absorption of a magnon, and we neglected the exception arising from the different normalization of the $\vk=0$ magnon mode (see App. \ref{app:a}). From the Fermi Golden rule we then obtain the spin current
\begin{align}
  j_x^{\rm s}(0) =&\, \frac{1}{\pi A S}
  \left( \sum_{\vp_{\parallel}} \frac{2 (S J_{\rm sd})^2 p_x^2}{V_0^2 (2 m V_0 - p_x^2)} \right)
  \left( \frac{1}{N}
  \sum_{\vk} \varepsilon_{\vk} b^0_{\vk} \right) \Delta T ,
\end{align}
consistent with Eq.\ (\ref{eq:Spgeneral}).

\bibliography{literature_lsse}

\end{document}